\documentstyle[12pt,epsf]{article}

\begin{document}

\baselineskip=18.8pt plus 0.2pt minus 0.1pt

\makeatletter

\renewcommand{\thefootnote}{\fnsymbol{footnote}}
\newcommand{\beq}{\begin{equation}}
\newcommand{\eeq}{\end{equation}}
\newcommand{\bea}{\begin{eqnarray}}
\newcommand{\eea}{\end{eqnarray}}
\newcommand{\nn}{\nonumber \\}
\newcommand{\hs}[1]{\hspace{#1}}
\newcommand{\vs}[1]{\vspace{#1}}
\newcommand{\Half}{\frac{1}{2}}
\newcommand{\p}{\partial}
\newcommand{\ol}{\overline}
\newcommand{\wt}[1]{\widetilde{#1}}
\newcommand{\ap}{\alpha'}
\newcommand{\bra}[1]{\left\langle  #1 \right\vert }
\newcommand{\ket}[1]{\left\vert #1 \right\rangle }
\newcommand{\vev}[1]{\left\langle  #1 \right\rangle }
\newcommand{\vac}{\ket{0}}
\newcommand{\ul}[1]{\underline{#1}}

\def\@bgnmark{<}
\def\@endmark{>}
\def\WKht{.85}
\def\WKsep{.4}
\def\WKrule{.03}
\newcount\@bgncnt
\newcount\@endcnt
\newcount\@h@ight
\newcount\TempCount
\newif\if@Exist
\newif\if@under
\newdimen\@tempdimc
\newdimen\@tempdimd
\newdimen\fixh@ight
\newdimen\h@ight 
\newdimen\w@dth  

\def\SEPbgn#1<#2#3#4\@@{\xdef\@MAE{#1}\xdef\@MARK{#2}
\xdef\@FRONT{#3}\xdef\@USIRO{#4}}
\def\SEPend#1>#2#3#4\@@{\xdef\@MAE{#1}\xdef\@MARK{#2}
\xdef\@FRONT{#3}\xdef\@USIRO{#4}}
\def\c@lc{
 \setbox0=\hbox{$\displaystyle \@FRONT$}
 \@tempdima\wd0 \if@under\@tempdimb\dp0\else\@tempdimb\ht0\fi
 \settowidth{\@tempdimc}{$\displaystyle \@MAE$}
 \settowidth{\@tempdimd}{$\displaystyle \@list$}
 \divide\@tempdima by2 \advance\@tempdima by \@tempdimc
 \advance\@tempdima by \@tempdimd}

\def\@dblfornoop#1\@@#2#3#4{}
\def\@dblfor#1;#2:=#3\do#4{\xdef\@fortmp{#3}\ifx\@fortmp\@empty \else%
 \expandafter\@dblforloop#3\@nil,\@nil,\@nil\@@#1#2{#4}\fi}
\def\@dblforloop#1,#2,#3\@@#4#5#6{\def#4{#1} \def#5{#2}%
 \ifx #4\@nnil \let\@nextwhile=\@dblfornoop \else%
 #6\relax \let\@nextwhile=\@dblforloop\fi\@nextwhile#3\@@#4#5{#6}}

\def\fin@endpt#1#2{
\@dblfor\MemBer;\NextmemBer:=#2\do{\def\@bject{#1}%
 \if \MemBer\@bject \xdef\@endpt{\NextmemBer} \@Existtrue\fi}}%
\def\fin@h@ight#1#2{
 \@tempcnta\z@%
 \@tfor\MEmber:=#2\do{\advance\@tempcnta\@ne%
 \ifnum \@tempcnta=#1 \@h@ight=\MEmber\fi}}

\def\wicksymbol#1#2#3#4#5{
 \@tempdima=#3 \advance\@tempdima-#1%
 \@tempdimc=#5\h@ight \@tempdimb=\@tempdimc\advance\@tempdimb-\w@dth%
 \@tempdimd=#2 \advance\@tempdimd-\fixh@ight
 \hskip#1%
 \if@under\def\tpdp{height}\def\tpht{depth}
 \else\def\tpdp{depth}\def\tpht{height}\fi
 \vrule \tpht\@tempdimc width\w@dth \tpdp-\@tempdimd \kern-\w@dth%
 \vrule \tpht\@tempdimc width\@tempdima \tpdp-\@tempdimb\kern-\w@dth%
 \@tempdimd=#4 \advance\@tempdimd-\fixh@ight
 \vrule \tpht \@tempdimc width\w@dth \tpdp-\@tempdimd}
\def\first#1{\expandafter\@mae#1\@nil}
\def\secnd#1{\expandafter\@ato#1\@nil}
\def\@mae#1;#2\@nil{#1}
\def\@ato#1;#2\@nil{#2}

\def\overwick{\wick[o]}
\def\underwick{\wick[u]}
\def\wick{\@ifnextchar[{\@wick}{\@wick[o]}}
\def\@wick[#1]#2#3{\def\b@dy{\displaystyle\@list}%
\@@wick{#1}{#2}{#3}}%
\def\Wwick#1#2#3#4{\def\b@dy{\wick[u]{#3}{#4}}%
\@@wick{o}{#1}{#2}}%
\def\@@wick#1#2#3{%
\if #1o\@underfalse\fixh@ight1.587ex\else\@undertrue\fixh@ight\z@\fi%
 \h@ight=\WKht ex \w@dth=\WKrule em%
 \@bgncnt\z@ \@endcnt\z@%
 \def\@list{} \def\bgnp@sition{} \def\endp@sition{}%
 \@ifundefined{DeclareOldFontCommand}{}{
 \def\rm{\mathrm} \def\bf{\mathbf} \def\cal{\mathcal}}
 \def\sqrt{\radical"270370}%
 \xdef\str@ng{#3}
 \@tfor\m@mber:=#3\do{%
 \ifx\m@mber\@bgnmark \advance\@bgncnt\@ne
  \expandafter\SEPbgn\str@ng\empty\@@ \c@lc
  \xdef\bgnp@sition{\bgnp@sition\@MARK,\the\@tempdima;\the\@tempdimb,}
  \xdef\@list{\@list{\@MAE}{\@FRONT}}
  \xdef\str@ng{\@USIRO}\fi
 \ifx \m@mber\@endmark \advance\@endcnt\@ne
  \expandafter\SEPend\str@ng\empty\@@ \c@lc
  \xdef\endp@sition{\endp@sition\@MARK,\the\@tempdima;\the\@tempdimb,}
  \xdef\@list{\@list{\@MAE}{\@FRONT}}
  \xdef\str@ng{\@USIRO}\fi}
  \xdef\@list{\@list\@USIRO}
 \ifnum\@bgncnt=\@endcnt \else%
 \@latexerr{The numbers of `<' and `>' do not match}%
 {You have written different numbers of < and >}\fi%
 \TempCount\z@ \@tfor\mmbr:=#2\do{\advance\TempCount\@ne}%
 \ifnum\@bgncnt=\TempCount \else%
 \@latexerr{The number of numbers in the first argument is different
 with that of contractions <...>}%
 {Give the same numbers of heights as the contractions <...>}\fi
 \mathop{\if@under\vtop\else\vbox\fi%
{\m@th\ialign{##\crcr%
\if@under
 \setbox0=\hbox{$\displaystyle\@list$}\dp0=\z@%
 \box0\crcr\noalign{\kern\WKsep ex\nointerlineskip}%
\else
 \noalign{\kern\WKsep ex}%
\fi
 $\m@th \TempCount\z@%
 \@dblfor\member;\nextmember:=\bgnp@sition\do{
 \advance\TempCount\@ne \xdef\@bgnpt{\nextmember}
 \@Existfalse%
 \fin@endpt{\member}{\endp@sition}%
 \if@Exist \else \@latexerr{The begin-mark `<\member' has no
corresponding end-mark `>\member'}{You should write coinciding label 
like <\member .. >\member}\fi%
 \fin@h@ight{\TempCount}{#2}%
 \setbox0=\hbox{%
 $\wicksymbol{\first\@bgnpt}{\secnd\@bgnpt}{\first\@endpt}%
 {\secnd\@endpt}{\@h@ight}$\hss}%
 \if@under\ht0\else\dp0\fi\z@\wd0\z@ \box0}$\crcr
\if@under
 \noalign{\kern\WKsep ex}%
\else
 \noalign{\kern\WKsep ex\nointerlineskip}%
 \setbox0=\hbox{$\b@dy$}\ht0=\fixh@ight%
 \box0\crcr
\fi
 }}}\limits}

\makeatother

\begin{titlepage}
\title{
\hfill\parbox{5cm}
{\normalsize CERN-PH-TH/2004-257\\{\tt hep-th/0412215}}\\
\vspace{1cm}
A Covariant Action with a Constraint and Feynman Rules for Fermions
in Open Superstring Field Theory}
\author{Yoji Michishita
\thanks{
{\tt Yoji.Michishita@cern.ch}
}
\\[7pt]
{\it Theory Division, CERN,
CH-1211, Geneva 23, Switzerland}}

\date{\normalsize December, 2004}
\maketitle
\thispagestyle{empty}

\begin{abstract}
\normalsize
In a way analogous to type IIB supergravity, we give a covariant action
for the fermion field supplemented with a constraint which should be 
imposed on equations of motion, in Berkovits' open superstring field theory.
From this action we construct Feynman rules for computing perturbative
amplitudes for fermions. We show that on-shell tree level 4-point amplitudes
computed by using these rules coincide with those of the first
quantization formalism.
\end{abstract}



\end{titlepage}

\section{Introduction}

In superstring theory supersymmetry plays crucial roles.
Therefore it is very important to consider the fermion sector as well as 
the boson sector. In this paper we consider this issue in the open 
superstring field theory proposed by Berkovits \cite{b96}. 
In \cite{b96} only the bosonic part was 
given, and some attempts to introduce fermions have been
made in \cite{b01}, where covariant equations of motion for the fermion field
have been given, and unfortunately it is impossible to write down an action 
from which those equations are derived. Some noncovariant actions have 
also been given in \cite{b01}. This has been achieved by introducing 
two or more additional string fields, 
but those string fields consist of both bosons and fermions, and  
the apparent forms of the bosonic parts of those actions are different from
that of \cite{b96}.

This situation is reminiscent of ordinary field theories with self-dual forms:
we can easily write down field equations, but naive attempts to construct
covariant actions from which those equations are derived fail. \cite{ms82}
(However introduction of an auxiliary field leads to a successful formulation.
\cite{pst}) One of this kind of theory is type IIB supergravity, which
has a 4-form with self-dual 5-form field strength. What we usually do in this
theory is as follows: we write down an 
action pretending that the 4-form has both the self-dual and the anti 
self-dual part,
and after we compute equations of motion for the self-dual field and others 
we impose the self-duality condition on them.
In this paper we will apply the same procedure in
the open superstring field theory. i.e. we introduce an additional string
field corresponding to the anti self-dual part, write down a covariant 
action, and
impose a constraint. We will show that the equations of motion derived from
our action reduce to those of \cite{b01} under the constraint. Then we will use
the action for deriving Feynman rules for computing perturbative
amplitudes for fermions. We show that 4-point on-shell tree level amplitudes
with fermions computed according to these rules coincide with those 
of the first quantization formalism.

\section{A covariant action for fermions with a constraint}

Let us recall type IIB supergravity. This theory has a 4-form, and its
field strength is self-dual. In general the kinetic term of a $(p-1)$-form
is given by 
$F_{\mu_1\mu_2\dots\mu_p}F^{\mu_1\mu_2\dots\mu_p}$, and when $p$ is half 
of spacetime dimension $D$ and $D=2$ mod 4, this is equal to
$2F^+_{\mu_1\mu_2\dots\mu_p}F^{-\,\mu_1\mu_2\dots\mu_p}$, where
$F^{\pm}_{\mu_1\mu_2\dots\mu_p}$ are the self-dual and the anti self-dual 
part of the field strength respectively. Since both parts appear in this form
of kinetic term, it does not extend to the case with only the self-dual part. 

We usually detour around this problem by writing down an action 
assuming temporarily
that the 5-form field strength has both self-dual and anti self-dual part, 
and imposing additional self-duality constraint after deriving equations of 
motion. The action for the metric, dilaton, NSNS $B$-field, and RR forms
$C_n$ is
\bea
S & = & \frac{1}{2\kappa^2}\int d^{10}x\sqrt{-g}\Bigg[
e^{-2\phi}\left(R+4g^{\mu\nu}\p_\mu\phi\p_\nu\phi
-\frac{1}{2}|H_3|^2\right) \nn
& & -\frac{1}{2}\left(
|F_1|^2+|\wt{F}_3|^2+\frac{1}{2}|\wt{F}_5|^2\right)\Bigg] \nn
& & -\frac{1}{4\kappa^2}\int C_4\wedge H_3\wedge F_3, \label{iibaction}
\eea
where $H$ and $F_{n+1}$ are field strengths of $B$ and $C_n$ respectively,
and
\bea
\wt{F}_3 & = & F_3-C_0\wedge H_3, \nn
\wt{F}_5 & = & F_5-\frac{1}{2}C_2\wedge H_3+\frac{1}{2}B_2\wedge F_3.
\eea
The self-duality condition is $*\wt{F}_5=\wt{F}_5$.
This is imposed on the solutions, and not on the action.
The equation of motion for $C_4$ derived from the above action is
$d*\wt{F}_5=H_3\wedge F_3$ and this is reduced to the Bianchi identity
under the self-duality condition.

Next we consider the fermion sector of Berkovits' open superstring field theory
\cite{b96} corresponding to one single BPS D-brane.
Extension to non-BPS D-branes or multiple D-brane case is
straightforward.
A natural string field $\Psi$ for fermions has $n_p=1/2$ and $n_g=0$, 
where $n_p$ and $n_g$ are picture number and ghost number respectively.
(For the assignment of these numbers see, for instance, \cite{bsz00}.)
This field corresponds to $\phi$-charge $-1/2$ vertex operators in the 
first quantization formalism.
At the linearized level this field should have the following gauge symmetry,
\beq
\delta\Psi=Q_B\Lambda_{1/2}+\eta_0\Lambda_{3/2},
\eeq
where $\Lambda_n$ are parameters with $(n_p,n_g)=(n,-1)$,
and the equation of motion should be $Q_B\eta_0\Psi=0$.

In the oscillator expression this field is expanded by the following states,
constructed by acting indicated oscillators so that they have indicated
$n_g$ and Grassmann parity:
\bea
\xi_0\{\beta_{n\leq -1},\gamma_{n\leq 0},b_{n\leq -1},c_{n\leq 0},
L^m_{n\leq -1},G^m_{n\leq -1}; n_g=0,\mbox{Grassmann even}\}
\ket{\Omega_{-1/2}^A}, \nn
\xi_0\{\beta_{n\leq -1},\gamma_{n\leq 0},b_{n\leq -1},c_{n\leq 0},
L^m_{n\leq -1},G^m_{n\leq -1}; n_g=0,\mbox{Grassmann odd}\}
\ket{\wt{\Omega}_{-1/2}^A}, \nn
\{\beta_{n\leq -2},\gamma_{n\leq 1},b_{n\leq -1},c_{n\leq 0},
L^m_{n\leq -1},G^m_{n\leq -1}; n_g=-1,\mbox{Grassmann even}\}
\ket{\Omega_{1/2}^A}, \nn
\{\beta_{n\leq -2},\gamma_{n\leq 1},b_{n\leq -1},c_{n\leq 0},
L^m_{n\leq -1},G^m_{n\leq -1}; n_g=-1,\mbox{Grassmann odd}\}
\ket{\wt{\Omega}_{1/2}^A},
\eea
where $\ket{\Omega_n^A}=c(0)e^{n\phi(0)}\Sigma^A(0)\ket{0}$ and 
$\ket{\wt{\Omega}_n^A}=c(0)e^{n\phi(0)}\wt{\Sigma}^A(0)\ket{0}$.
$\Sigma^A$ and $\wt{\Sigma}^A$ are spin operators with positive and negative
chirality respectively.
Note that on $\ket{\Omega_{\pm 1/2}^A}$ and 
$\ket{\wt{\Omega}_{\pm 1/2}^A}$,
$\beta$, $\gamma$ and $G^m$ have integer mode numbers. 
The operators $e^{-\frac{1}{2}\phi}\Sigma^A$ and 
$e^{-\frac{1}{2}\phi}\wt{\Sigma}^A$ should be regarded as
Grassmann odd and even respectively, then all the above states are 
Grassmann odd. We set the coefficients of these states 
Grassmann odd, so that they represent fermions. Then $\Psi$ is
Grassmann even.

Here are low lying states for the expansion of $\Psi$ in the flat background.
\begin{center}
\begin{tabular}{|c|r|r|r|r|}\hline
 level & states with $\xi_0$ & states with $\xi_0$ 
 & states without $\xi_0$ & states without $\xi_0$ \\
 $(L_0-\ap k^2)$ & and without $c_0$ & and with $c_0$
 & and without $c_0$ & and with $c_0$ \\ \hline
 0 & $\xi_0\ket{\Omega_{-1/2}^A,k}$ & none 
   & $b_{-1}\ket{\wt{\Omega}_{1/2}^A,k}$ & none \\ \hline
 1 & $\xi_0\beta_{-1}\gamma_0\ket{\Omega_{-1/2}^A,k}$
   & $\xi_0c_0\beta_{-1}\ket{\wt{\Omega}_{-1/2}^A,k}$
   & $b_{-2}\ket{\wt{\Omega}_{1/2}^A,k}$ & none \\
   & $\xi_0b_{-1}\gamma_0\ket{\wt{\Omega}_{-1/2}^A,k}$
   & $\xi_0c_0b_{-1}\ket{\Omega_{-1/2}^A,k}$
   & $\beta_{-2}\ket{\Omega_{1/2}^A,k}$ & \\
   & $\xi_0\alpha_{-1}^\mu\ket{\Omega_{-1/2}^A,k}$ &
   & $b_{-1}\alpha^\mu_{-1}\ket{\wt{\Omega}_{1/2}^A,k}$ & \\
   & $\xi_0\psi_{-1}^\mu\ket{\wt{\Omega}_{-1/2}^A,k}$ &
   & $b_{-1}\psi^\mu_{-1}\ket{\Omega_{1/2}^A,k}$ & \\
   &
   &
   & $b_{-2}b_{-1}\gamma_1\ket{\Omega_{1/2}^A,k}$ & \\
   &
   &
   & $\beta_{-2}b_{-1}\gamma_1\ket{\wt{\Omega}_{1/2}^A,k}$ & \\ \hline
\end{tabular}
\end{center}
In the above table 
$\ket{\Omega_n^A,k}=c(0)e^{n\phi(0)}\Sigma^A(0)e^{ik\cdot X(0)}\ket{0}$
and $\ket{\wt{\Omega}_n^A,k}
=c(0)e^{n\phi(0)}\wt{\Sigma}^A(0)e^{ik\cdot X(0)}\ket{0}$.

Naively kinetic term of $\Psi$ is $\vev{\vev{(Q_B\Psi)(\eta_0\Psi)}}$, 
but this vanishes because of the picture number conservation law.
One may think we can introduce
picture changing operators to give correct kinetic term, but it is well 
known that this causes divergent contact term problems \cite{w89}, 
and modifies the equation of motion.

Thus it seems impossible to construct a consistent kinetic 
term for $\Psi$.
However, as has been done in \cite{b01}, we can construct a nonlinear
extension of equations of motion and gauge symmetry:
\bea
\eta_0(G^{-1}(Q_BG)) & = & -(\eta_0\Psi)^2, \label{coveqb} \\
Q_B(G(\eta_0\Psi)G^{-1}) & = & 0, \label{coveqf}
\eea
\bea
\delta G & = & G(\eta_0\Lambda_1-\{\eta_0\Psi,\Lambda_{1/2}\})+(Q\Lambda_0)G,
\label{covsymb} \\
\delta\Psi & = & \eta_0\Lambda_{3/2}+[\Psi,\eta_0\Lambda_1]+Q\Lambda_{1/2}
+\{G^{-1}(Q_BG),\Lambda_{1/2}\}, \label{covsymf}
\eea
where $G=e^\Phi$, and $\Phi$ is the string field for bosons.

Comparing this with ordinary field theories with self-dual forms, we 
notice that we are in a similar situation: We have equations of motion, 
but cannot write down an covariant action, in particular kinetic term, which 
reproduces them. Then it is natural to think about doing the same thing
as in type IIB supergravity. i.e. adding an additional field corresponding to
the anti self-dual part, writing down an action, and impose a constraint 
corresponding to the self-duality condition. Let us call the additional string
field $\Xi$, and we infer the action at the linearized level is 
given by the product of $\Xi$ and $\Psi$ just as the kinetic terms of 
forms are given by the product of the self-dual and the anti self-dual part:
\beq
S_F=-\frac{1}{2g^2}\vev{\vev{(Q_B\Xi)(\eta_0\Psi)}}.
\eeq
From this we can see that $\Xi$ has $(n_p,n_g)=(-1/2,0)$, and is 
Grassmann even. 
In the oscillator expression $\Xi$ is expanded by the following states:
\bea
\xi_0\{\beta_{n\leq 0},\gamma_{n\leq -1},b_{n\leq -1},c_{n\leq 0},
L^m_{n\leq -1},G^m_{n\leq -1}; n_g=0,\mbox{Grassmann odd}\}
\ket{\Omega_{-3/2}^A}, \nn
\xi_0\{\beta_{n\leq 0},\gamma_{n\leq -1},b_{n\leq -1},c_{n\leq 0},
L^m_{n\leq -1},G^m_{n\leq -1}; n_g=0,\mbox{Grassmann even}\}
\ket{\wt{\Omega}_{-3/2}^A}, \nn
\{\beta_{n\leq -1},\gamma_{n\leq 0},b_{n\leq -1},c_{n\leq 0},
L^m_{n\leq -1},G^m_{n\leq -1}; n_g=-1,\mbox{Grassmann odd}\}
\ket{\Omega_{-1/2}^A}, \nn
\{\beta_{n\leq -1},\gamma_{n\leq 0},b_{n\leq -1},c_{n\leq 0},
L^m_{n\leq -1},G^m_{n\leq -1}; n_g=-1,\mbox{Grassmann even}\}
\ket{\wt{\Omega}_{-1/2}^A}.
\eea

Here are low lying states for the expansion of $\Xi$ in the flat background.
\begin{center}
\begin{tabular}{|c|r|r|r|r|}\hline
 level & states with $\xi_0$ & states with $\xi_0$ 
 & states without $\xi_0$ & states without $\xi_0$ \\
 $(L_0-\ap k^2)$ & and without $c_0$ & and with $c_0$
 & and without $c_0$ & and with $c_0$ \\ \hline
 0 & $\xi_0\ket{\wt{\Omega}_{-3/2}^A,k}$
   & $\xi_0c_0\beta_0\ket{\Omega_{-3/2}^A,k}$  & none & none \\ \hline
 1 & $\xi_0\beta_0\gamma_{-1}\ket{\wt{\Omega}_{-3/2}^A,k}$
   & $\xi_0c_0\beta_{-1}\ket{\Omega_{-3/2}^A,k}$
   & $\beta_{-1}\ket{\wt{\Omega}_{-1/2}^A,k}$ & none \\
   & $\xi_0\beta_0c_{-1}\ket{\Omega_{-3/2}^A,k}$
   & $\xi_0c_0b_{-1}\ket{\wt{\Omega}_{-3/2}^A,k}$
   & $b_{-1}\ket{\Omega_{-1/2}^A,k}$ & \\
   & $\xi_0\alpha_{-1}^\mu\ket{\wt{\Omega}_{-3/2}^A,k}$
   & $\xi_0c_0\beta_0\alpha_{-1}^\mu\ket{\Omega_{-3/2}^A,k}$ 
   & & \\
   & $\xi_0\psi_{-1}^\mu\ket{\Omega_{-3/2}^A,k}$
   & $\xi_0c_0\beta_0\psi_{-1}^\mu\ket{\wt{\Omega}_{-3/2}^A,k}$ 
   & & \\
   & 
   & $\xi_0c_0(\beta_0)^2\gamma_{-1}\ket{\Omega_{-3/2}^A,k}$ 
   & & \\
   & 
   & $\xi_0c_0(\beta_0)^2c_{-1}\ket{\wt{\Omega}_{-3/2}^A,k}$
   & & \\ \hline
\end{tabular}
\end{center}

The equations of motion for $\Psi$ and $\Xi$ are
\bea
Q_B\eta_0\Xi  & = & 0, \\
Q_B\eta_0\Psi & = & 0.
\eea
We have to put the following constraint to eliminate the superfluous degrees of 
freedom and to make the equation of motion of $\Psi$ or $\Xi$ trivial. 
\bea
Q_B\Xi=\eta_0\Psi.
\eea
This condition means that the ``self-dual'' and ``anti self-dual'' part
correspond to $\frac{1}{2}(Q_B\Xi\pm\eta_0\Psi)$,
rather than $\Psi$ and $\Xi$.

The above action is easily extended to a nonlinear interacting system:
\bea
S & = & S_B+S_F \label{covaction}, \\
S_B & = & \frac{1}{2g^2}\vev{\vev{G^{-1}(Q_BG)G^{-1}(\eta_0G)
 -\int_0^1dt G_t^{-1}(\p_tG_t)\{G_t^{-1}(Q_BG_t),G_t^{-1}(\eta_0G_t)\} }}, \\
S_F & = & -\frac{1}{2g^2}\vev{\vev{(Q_B\Xi)G(\eta_0\Psi)G^{-1}}},
\eea
where $G_t=e^{t\Phi}$. $S_B$ is the bosonic part given in \cite{b96}.

The equations of motion for $\Phi$, $\Psi$ and $\Xi$ are
\bea
\eta_0(G^{-1}(Q_BG)) & = & -\frac{1}{2}(\eta_0\Psi)G^{-1}(Q_B\Xi)G 
 -\frac{1}{2}G^{-1}(Q_B\Xi)G(\eta_0\Psi), \\
\eta_0(G^{-1}(Q_B\Xi)G) & = & 0, \\
Q_B(G(\eta_0\Psi)G^{-1}) & = & 0.
\eea
The constraint is extended to 
\beq
Q_B\Xi=G(\eta_0\Psi)G^{-1}.
\eeq
Under this constraint either of the equations of motion for $\Psi$ and $\Xi$
can be regarded as trivial, and the three equations of motion are reduced to 
eq.(\ref{coveqb}) and eq.(\ref{coveqf}).

The action (\ref{covaction}) has the following gauge symmetry.
\bea
\delta G & = & G(\eta_0\Lambda_1)+(Q_B\Lambda_0)G, \\
\delta\Psi & = & \eta_0\Lambda_{3/2} +[\Psi,\eta_0\Lambda_1], \\
\delta\Xi  & = &   Q_B \Lambda_{-1/2}+[Q_B\Lambda_0,\Xi].
\eea
This symmetry is consistent with the constraint:
\bea
\delta(Q_B\Xi-G\eta_0\Psi G^{-1}) & = & [Q_B\Lambda_0,
 Q_B\Xi-G(\eta_0\Psi)G^{-1}].
\eea
However eq.(\ref{coveqb}) and eq.(\ref{coveqf}) have larger gauge symmetry:
the transformations of $G$ and $\Psi$ have an extra parameter $\Lambda_{1/2}$.
The action (\ref{covaction}) does not have this symmetry. 
Thus we have an enhanced symmetry when we impose the constraint. 
Again this is similar to type IIB supergravity:
(Fermionic extension of) the action (\ref{iibaction}) 
does not have local supersymmetry,
but under the self-duality constraint the equations of motion do.

\section{Feynman rules and tree level 4-point amplitudes}

One of the advantages of having an action, even though it must be supplemented
with the constraint, is that we can construct Feynman rules 
for computing perturbative amplitudes. 
This is somewhat similar to what has been done for self-dual fields 
in \cite{aw83}.

First let us expand $S_F$
\bea
S_F & = & -\frac{1}{2g^2}\Big\langle\Big\langle(Q_B\Xi)(\eta_0\Psi)
-\{(Q_B\Xi)(\eta_0\Psi)+(\eta_0\Psi)(Q_B\Xi)\}\Phi \nn
& & 
-\frac{1}{2}\{(Q_B\Xi)(\eta_0\Psi)-(\eta_0\Psi)(Q_B\Xi)\}\Phi\Phi
+(Q_B\Xi)\Phi(\eta_0\Psi)\Phi \nn
& & 
+\dots\Big\rangle\Big\rangle \\
& = & -\frac{1}{2g^2}\sum_{n\geq 0,m\geq 0}\frac{(-1)^m}{n!m!}
\vev{\vev{(Q_B\Xi)\Phi^n(\eta_0\Psi)\Phi^m}}.
\eea
From cubic and higher terms of this expansion we can read off 
interaction vertices. Since the ``anti self-dual'' part should decouple, 
we project out the component which does not satisfy the linearized 
constraint $Q_B\Xi=\eta_0\Psi$.
i.e. $Q_B\Xi$ and $\eta_0\Psi$ in these vertices should be replaced by 
$\omega\equiv\frac{1}{2}(Q_B\Xi+\eta_0\Psi)$.
Then we can see that only those with odd $\Phi$s survive:

For even $N$,
\bea
& &
-\frac{1}{2g^2}\sum_{n\geq 0,m\geq 0,n+m=N}\frac{(-1)^m}{n!m!}
\vev{\vev{\omega\Phi^n\omega\Phi^m}} \nn
& & =
-\frac{1}{4g^2}\sum_{n\geq 0,m\geq 0,n+m=N}
\frac{1}{n!m!}((-1)^m-(-1)^n)
\vev{\vev{\omega\Phi^n\omega\Phi^m}} \nn
& & = 0.
\eea

For odd $N$,
\bea
& &
-\frac{1}{2g^2}\sum_{n\geq 0,m\geq 0,n+m=N}\frac{(-1)^m}{n!m!}
\vev{\vev{\omega\Phi^n\omega\Phi^m}} \nn
& & =
-\frac{1}{2g^2}\sum_{m>n\geq 0,n+m=N}
\frac{1}{n!m!}((-1)^m-(-1)^n)
\vev{\vev{\omega\Phi^n\omega\Phi^m}} \nn
& & =
-\frac{1}{g^2}\sum_{m>n\geq 0,n+m=N}\frac{(-1)^m}{n!m!}
\vev{\vev{\omega\Phi^n\omega\Phi^m}}.
\eea

If $\omega$ is connected to an external leg, we can safely replace it
by $\eta_0\Psi$. Then we can see easily that the 3-point vertex 
reproduces 3-point tree level amplitudes in the first quantization formalism.

We have to give the propagators for $\Xi$ and $\Psi$ to 
complete the Feynman rules.
First let us recall the propagator for $\Phi$.\cite{be01}
A convenient gauge fixing condition for the linearized gauge transformation
is $G^-_0\Phi=\wt{G}^-_0\Phi=0$, where $G^-_0=b_0$, 
$\wt{G}^-_0=\{Q_B,J^{--}_0\}$ and $J^{--}_0=\oint\frac{dz}{2\pi i}b(z)\xi(z)$.
Under this condition the propagator $P\equiv\underwick{1}{<1\Phi>1\Phi}$
is given by $P=(L^{\rm tot}_0)^{-2}G^-_0\wt{G}^-_0$. $L^{\rm tot}$ is the 
total Virasoro operator.
For $\Xi$ and $\Psi$, since the same gauge fixing condition cannot be 
imposed on 
the action, the kinetic term in the action cannot be used directly to 
compute the propagator under this gauge fixing condition, 
but it helps us to guess the correct form of the propagators:
Only $\underwick{1}{<1\Xi>1\Psi}$ and $\underwick{1}{<1\Psi>1\Xi}$
are nonzero.
Since propagating degrees of freedom should satisfy the constraint, 
we can think the same gauge fixing condition can be effectively
imposed on $\Xi$ and $\Psi$, and therefore the propagator is given by 
the same one as $\Phi$:
$\underwick{1}{<1\Xi>1\Psi}=\underwick{1}{<1\Psi>1\Xi}=-2P$.
The factor $-2$ comes from the difference of the coefficients of the kinetic 
terms in $S_B$ and $S_F$.

Strictly speaking, these propagators are given by
\bea
\underwick{1}{<1\Psi>1\Xi} & = & -2\sum_i\ket{i,(1/2,0)}\bra{i,(-3/2,2)}'
 (L^{\rm tot}_0)^{-2}G^-_0\wt{G}^-_0\sum_j
 \ket{j,(-1/2,2)}'\bra{j,(-1/2,0)}, \nn
\underwick{1}{<1\Xi>1\Psi} & = & -2\sum_i\ket{i,(-1/2,0)}\bra{i,(-1/2,2)}'
 (L^{\rm tot}_0)^{-2}G^-_0\wt{G}^-_0\sum_j
 \ket{j,(-3/2,2)}'\bra{j,(1/2,0)},
\eea
where $\{\ket{i,(n,m)}\}$ are bases for $(n_p,n_g)=(n,m)$ states satisfying 
the gauge fixing condition, and $\{\bra{i,(n,m)}'\}$ are conjugate bases
satisfying $\bra{i,(-n-1,-m+2)}'\ket{j,(n,m)}=\delta_{ij}$.
The strips corresponding to these propagators have the Ramond boundary 
condition.
Since we set $\ket{i,n}$ Grassmann odd, we have an additional minus sign
for each fermion loop.

Practically we need the propagator between $\omega$s:
\beq
\underwick{1}{<1\omega>1\omega}
=\frac{1}{4}(\underwick{1}{(Q_B<1\Xi)(\eta_0>1\Psi)}
+\underwick{1}{(\eta_0<1\Psi)(Q_B>1\Xi)}).
\eeq

\begin{figure}[htdp]
\begin{center}
\leavevmode
\epsfbox{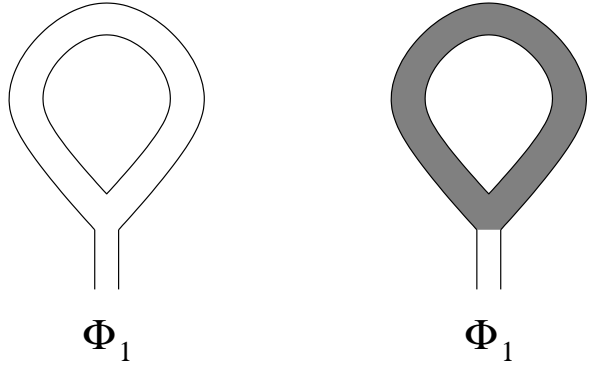}
\caption{1-loop tadpole diagrams}
\label{tad}
\end{center}
\end{figure}

As an immediate check of these rules, let us calculate 1-loop tadpoles.
As is indicated in Fig.\ref{tad}, 
we have bosonic loop and fermionic loop contributions.
We show fermions by shaded strips, and bosons by unshaded strips.
Noting that the bosonic 3-point vertex is given by 
$-\frac{1}{6}\vev{\vev{\Phi\{(Q_B\Phi)(\eta_0\Phi)+(\eta_0\Phi)(Q_B\Phi)\}}}$,
the contribution of the bosonic loop is
\beq
\left(-\frac{1}{6}\right)\cdot 3 \cdot\vev{\vev{\{
(\underwick{1}{(Q_B<1\Phi)(\eta_0>1\Phi)}
+\underwick{1}{(\eta_0<1\Phi)(Q_B>1\Phi)}\}\Phi_1}}.
\eeq
The contribution of the fermion loop is
\bea
(-1)\cdot\frac{1}{4}\cdot\vev{\vev{\{
(\underwick{1}{Q_B<1\Psi)(\eta_0>1\Xi)}
+\underwick{1}{(\eta_0<1\Xi)(Q_B>1\Psi)}\}\Phi_1}}.
\eea
By $\underwick{1}{<1\Xi>1\Psi}=\underwick{1}{<1\Psi>1\Xi}=
-2\underwick{1}{<1\Phi>1\Phi}$, we can see the above two contributions 
are in the same form. The differences are the signs and 
the boundary conditions on the boundary of the loops.
This coincides with the expected result.

Next let us calculate on-shell 4-point tree level amplitudes with fermions.
In \cite{be01} it has been shown that the bosonic part of the superstring
field theory reproduces tree level on-shell 4-boson amplitude in the first 
quantization formalism. Therefore we expect that our Feynman rules reproduce
fermion amplitudes. 
Since we take external legs on-shell, they satisfy the linearized equations of 
motion $Q_B\eta_0\Phi=Q_B\eta_0\Psi=0$. In the following we mostly follow
the notation of \cite{be01}, and we do not explain each step of the 
calculation in detail here, because much of the details is parallel 
to the argument in \cite{be01}.

First we calculate the 4-fermion amplitude $A_{FFFF}$.
We sum up those with 4 external fermion legs in the order of 
$\Psi_4\Psi_1\Psi_2\Psi_3$ and its cyclic permutations,
and compare with the corresponding one in the first quantization formalism.
Those in other orders can be considered similarly.
\begin{figure}[htdp]
\begin{center}
\leavevmode
\epsfbox{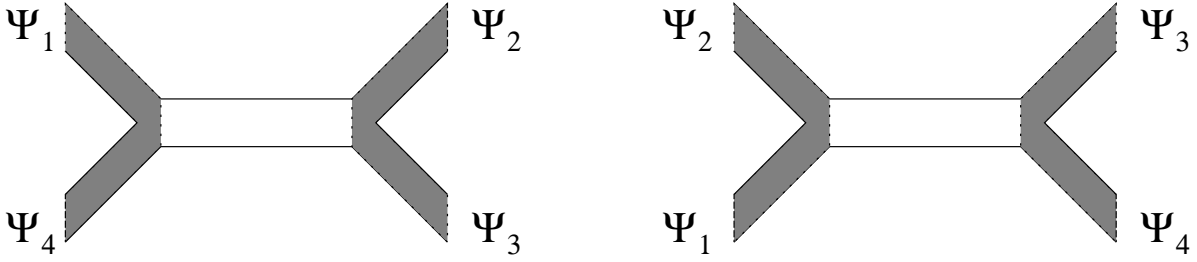}
\caption{4-fermion interaction}
\label{ffff}
\end{center}
\end{figure}
Since we have no 4-point vertex with fermions, our task is
to compute ``s-channel'' contribution $A_{FFFF}^s$ and 
``t-channel'' contribution $A_{FFFF}^t$ indicated in Fig.\ref{ffff}.
In the 4-boson case the 4-boson vertex played a crucial role
when we combine s- and t-channel contributions into one single integral.
In the present case we expect that $A_{FFFF}^s+A_{FFFF}^t$ itself is
expressed by one single integral. Let us see if this is the case.

The s-channel contribution is
\bea
A_{FFFF}^s
 & = & g^{-2}\underwick{1}{
  \langle\langle(\eta_0\Psi_4)(\eta_0\Psi_1)<1\Phi\rangle\rangle
  \langle\langle >1\Phi(\eta_0\Psi_2)(\eta_0\Psi_3)\rangle\rangle} \nn
 & = & g^{-2}
 \vev{(\eta_0\Psi_4)(\eta_0\Psi_1)P(\eta_0\Psi_2)(\eta_0\Psi_3)}_W.
\eea
Then we deform off the contour of $Q_B=\oint\frac{dz}{2\pi i}j_B(z)$
in $P=(L^{\rm tot}_0)^{-2}G^-_0\wt{G}^-_0
=(L^{\rm tot}_0)^{-2}G^-_0\{Q_B,J^{--}_0\}$ away from $J^{--}(z)$, 
effectively replacing $P$ by $(L^{\rm tot}_0)^{-1}J^{--}_0$:
\bea
A_{FFFF}^s
 & = & g^{-2}\vev{(\eta_0\Psi_4)(\eta_0\Psi_1)
  (L^{\rm tot}_0)^{-1}J^{--}_0(\eta_0\Psi_2)(\eta_0\Psi_3)}_W \nn
 & = & g^{-2}\int_0^\infty d\tau
  \vev{\int_c \frac{dw}{2\pi i}J^{--}(w)(\eta_0\Psi_4)(\eta_0\Psi_1)
  (\eta_0\Psi_2)(\eta_0\Psi_3)}_W \nn
 & = & -g^{-2}\int_0^\delta d\alpha\frac{d\tau}{d\alpha}
  \vev{\int_{\bar{c}}\frac{dz}{2\pi i}\frac{dz}{dw}
  J^{--}(z)\eta_0\Psi_4(-\alpha^{-1})\eta_0\Psi_1(-\alpha)
  \eta_0\Psi_2(\alpha)\eta_0\Psi_3(\alpha^{-1})}.
\eea
Readers can find the definitions of $\tau$, $\alpha$, $\delta$, $c$,
$\bar{c}$, $w=w(z)$ and $W$ in \cite{be01}.

Similarly,
\bea
A_{FFFF}^t
 & = & g^{-2}\underwick{1}{
  \langle\langle(\eta_0\Psi_3)(\eta_0\Psi_4)<1\Phi\rangle\rangle
  \langle\langle >1\Phi(\eta_0\Psi_1)(\eta_0\Psi_2)\rangle\rangle} \nn
 & = & g^{-2}\int_\delta^1 d\alpha\frac{d\tau}{d\alpha}
  \vev{\int_{\bar{c}}\frac{dz}{2\pi i}\frac{dz}{dw}
  J^{--}(z)\eta_0\Psi_3(\alpha^{-1})\eta_0\Psi_4(-\alpha^{-1})
  \eta_0\Psi_1(-\alpha)\eta_0\Psi_2(\alpha)}.
\eea
We see that the sum of $A_{FFFF}^s$ and $A_{FFFF}^t$ is one single integral
over the moduli space, and coincides with the amplitude in the first 
quantization formalism:
\bea
A_{FFFF} & = & -g^{-2}\int_0^1 d\alpha\frac{d\tau}{d\alpha}
  \vev{\int_{\bar{c}}\frac{dz}{2\pi i}\frac{dz}{dw}
  J^{--}(z)\eta_0\Psi_4(-\alpha^{-1})\eta_0\Psi_1(-\alpha)
  \eta_0\Psi_2(\alpha)\eta_0\Psi_3(\alpha^{-1})} \nn
& = & -g^{-2}\int_0^1 d\alpha
  \vev{\int d^2z\mu_\alpha(z,\bar{z})J^{--}(z)
  \eta_0\Psi_4(-\alpha^{-1})\eta_0\Psi_1(-\alpha)
  \eta_0\Psi_2(\alpha)\eta_0\Psi_3(\alpha^{-1})},
\eea
where $\mu_\alpha(z,\bar{z})$ is the Beltrami differential corresponding to
the $\alpha$ modulus.

\begin{figure}[htdp]
\begin{center}
\leavevmode
\epsfbox{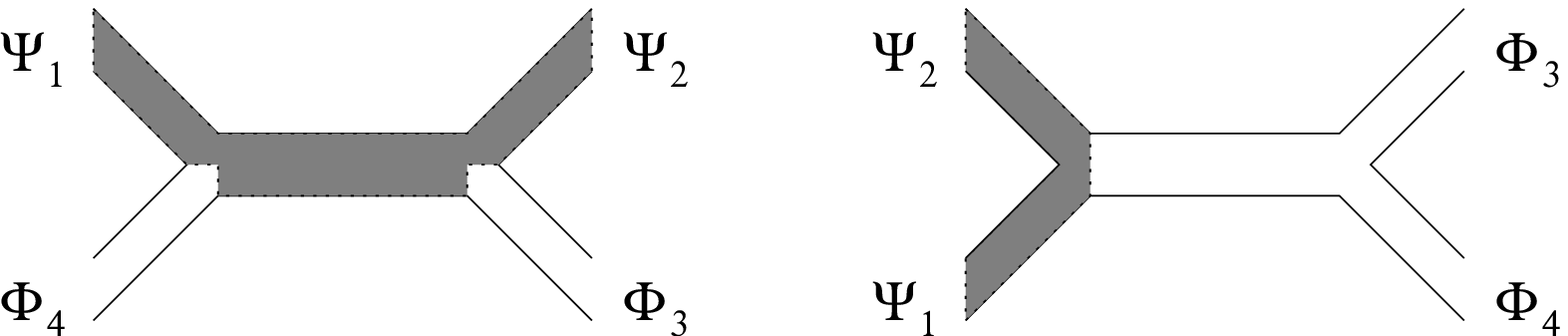}
\caption{2-boson 2-fermion interaction in the 
order of $\Phi_4\Psi_1\Psi_2\Phi_3$}
\label{ffbb}
\end{center}
\end{figure}

The second example is the 2-boson 2-fermion amplitude $A_{FFBB}$ in the 
order of $\Phi_4\Psi_1\Psi_2\Phi_3$, indicated in Fig.\ref{ffbb}.
The ``s-channel'' contribution is 
\bea
A_{FFBB}^s
 & = & g^{-2}\underwick{1}{
  \langle\langle\Phi_4(\eta_0\Psi_1)<1\omega\rangle\rangle
  \langle\langle >1\omega(\eta_0\Psi_2)\Phi_3\rangle\rangle} \nn
 & = & -\frac{1}{2}g^{-2}\Big[
  \vev{(Q_B\Phi_4)(\eta_0\Psi_1)P(\eta_0\Psi_2)(\eta_0\Phi_3)}_W
 +\vev{(\eta_0\Phi_4)(\eta_0\Psi_1)P(\eta_0\Psi_2)(Q_B\Phi_3)}_W
 \Big] \nn
 & = & \frac{1}{2}g^{-2}\int_0^\delta d\alpha\frac{d\tau}{d\alpha}
  \Bigg\langle\int_{\bar{c}}\frac{dz}{2\pi i}\frac{dz}{dw}
  J^{--}(z)\Big[Q_B\Phi_4(-\alpha^{-1})\eta_0\Psi_1(-\alpha)
  \eta_0\Psi_2(\alpha)\eta_0\Phi_3(\alpha^{-1}) \nn
 & & +\eta_0\Phi_4(-\alpha^{-1})\eta_0\Psi_1(-\alpha)
  \eta_0\Psi_2(\alpha)Q_B\Phi_3(\alpha^{-1})\Big]\Bigg\rangle.
\eea
In the second line we deformed off contours of 
$Q_B=\oint\frac{dz}{2\pi i}j_B(z)$ and $\eta_0=\oint\frac{dz}{2\pi i}\eta(z)$
away from $\Psi$ and $\Xi$ in the propagators to other fields.

Similarly the ``t-channel'' contribution is
\bea
A_{FFBB}^t
 & = & 3\cdot\left(-\frac{1}{6}\right)g^{-2}\underwick{1}{
  \langle\langle(\eta_0\Psi_1)(\eta_0\Psi_2)<1\Phi\rangle\rangle
  \langle\langle >1\Phi\{(Q_B\Phi_3)(\eta_0\Phi_4)
   +(\eta_0\Phi_3)(Q_B\Phi_4)\}\rangle\rangle} \nn
 & = & -\frac{1}{2}g^{-2}
 \vev{(\eta_0\Psi_1)(\eta_0\Psi_2)P
  \{(Q_B\Phi_3)(\eta_0\Phi_4)+(\eta_0\Phi_3)(Q_B\Phi_4)\}}_W
 \nn
 & = & \frac{1}{2}g^{-2}\int_\delta^1 d\alpha\frac{d\tau}{d\alpha}
  \Bigg\langle\int_{\bar{c}}\frac{dz}{2\pi i}\frac{dz}{dw}
  J^{--}(z)\Big[Q_B\Phi_4(-\alpha^{-1})\eta_0\Psi_1(-\alpha)
  \eta_0\Psi_2(\alpha)\eta_0\Phi_3(\alpha^{-1}) \nn
 & & +\eta_0\Phi_4(-\alpha^{-1})\eta_0\Psi_1(-\alpha)
  \eta_0\Psi_2(\alpha)Q_B\Phi_3(\alpha^{-1})\Big]\Bigg\rangle.
\eea
Again the sum of these contributions is expressed by one single integral
and gives the amplitude of the first quantization formalism:
\bea
A_{FFBB}
 & = & \frac{1}{2}g^{-2}\int_0^1 d\alpha\frac{d\tau}{d\alpha}
  \Bigg\langle\int_{\bar{c}}\frac{dz}{2\pi i}\frac{dz}{dw}
  J^{--}(z)\Big[Q_B\Phi_4(-\alpha^{-1})\eta_0\Psi_1(-\alpha)
  \eta_0\Psi_2(\alpha)\eta_0\Phi_3(\alpha^{-1}) \nn
 & & +\eta_0\Phi_4(-\alpha^{-1})\eta_0\Psi_1(-\alpha)
  \eta_0\Psi_2(\alpha)Q_B\Phi_3(\alpha^{-1})\Big]\Bigg\rangle \nn
 & = & g^{-2}\int_0^1 d\alpha\frac{d\tau}{d\alpha}
  \Bigg\langle\int_{\bar{c}}\frac{dz}{2\pi i}\frac{dz}{dw}
  J^{--}(z)Q_B\Phi_4(-\alpha^{-1})\eta_0\Psi_1(-\alpha)
  \eta_0\Psi_2(\alpha)\eta_0\Phi_3(\alpha^{-1})\Bigg\rangle \nn
 & = & 
  g^{-2}\int_0^1 d\alpha\Bigg\langle
  \int d^2z\mu_\alpha(z,\bar{z})J^{--}(z)
  Q_B\Phi_4(-\alpha^{-1})\eta_0\Psi_1(-\alpha)
  \eta_0\Psi_2(\alpha)\eta_0\Phi_3(\alpha^{-1})\Bigg\rangle.
\eea
In the second equality we exchanged $Q_B$ and $\eta_0$ on 
$\Phi_3$ and $\Phi_4$ in the second term of the first equality
by contour deformation. This manipulation leaves a total 
derivative term with respect to $\alpha$, but by the canceled propagator 
argument we can drop it.

\begin{figure}[htdp]
\begin{center}
\leavevmode
\epsfbox{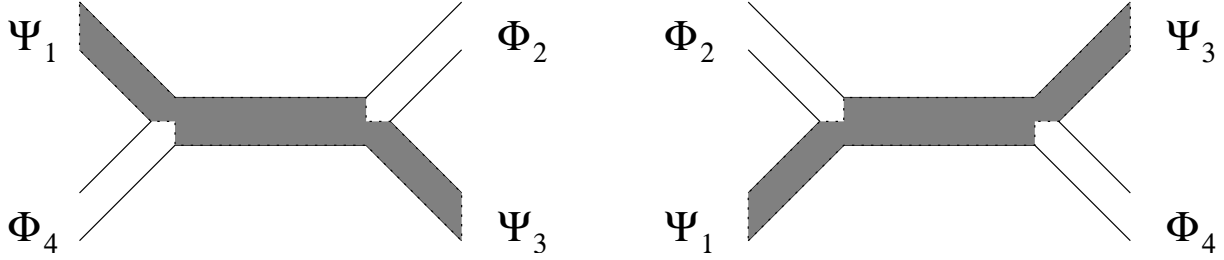}
\caption{2-boson 2-fermion interaction in the 
order of $\Psi_1\Phi_2\Psi_3\Phi_4$}
\label{fbfb}
\end{center}
\end{figure}

Finally we compute the 2-boson 2-fermion amplitude $A_{FBFB}$ in the 
order of $\Psi_1\Phi_2\Psi_3\Phi_4$ indicated in Fig.\ref{fbfb}.
\bea
A_{FBFB}^s
 & = & -g^{-2}\underwick{1}{
  \langle\langle\Phi_4(\eta_0\Psi_1)<1\omega\rangle\rangle
  \langle\langle >1\omega\Phi_2(\eta_0\Psi_3)\rangle\rangle} \nn
 & = & \frac{1}{2}g^{-2}[
  \vev{(Q_B\Phi_4)(\eta_0\Psi_1)P(\eta_0\Phi_2)(\eta_0\Psi_3)}_W
 +\vev{(\eta_0\Phi_4)(\eta_0\Psi_1)P(Q_B\Psi_2)(\eta_0\Psi_3)}_W ] \nn
 & = & -\frac{1}{2}g^{-2}\int_0^\delta d\alpha\frac{d\tau}{d\alpha}
  \Bigg\langle\int_{\bar{c}}\frac{dz}{2\pi i}\frac{dz}{dw}
  J^{--}(z)\Big[
 Q_B\Phi_4(-\alpha^{-1})\eta_0\Psi_1(-\alpha)
  \eta_0\Phi_2(\alpha)\eta_0\Psi_3(\alpha^{-1}) \nn
 & & +\eta_0\Phi_4(-\alpha^{-1})\eta_0\Psi_1(-\alpha)
  Q_B\Phi_2(\alpha)\eta_0\Psi_3(\alpha^{-1})
\Big]\Bigg\rangle.
\eea
\bea
A_{FBFB}^t
 & = & -g^{-2}\underwick{1}{
  \langle\langle(\eta_0\Psi_1)\Phi_2<1\omega\rangle\rangle
  \langle\langle >1\omega(\eta_0\Psi_3)\Phi_4\rangle\rangle} \nn
 & = & \frac{1}{2}g^{-2}[
 \vev{(\eta_0\Psi_1)(\eta_0\Phi_2)P(\eta_0\Psi_3)(Q_B\Phi_4)}
 +\vev{(\eta_0\Psi_1)(Q_B\Phi_2)P(\eta_0\Psi_3)(\eta_0\Phi_4)}
 ] \nn
 & = &
  -\frac{1}{2}g^{-2}\int_\delta^1 d\alpha\frac{d\tau}{d\alpha}
  \Bigg\langle\int_{\bar{c}}\frac{dz}{2\pi i}\frac{dz}{dw}
  J^{--}(z)\Big[
 Q_B\Phi_4(-\alpha^{-1})\eta_0\Psi_1(-\alpha)
  \eta_0\Phi_2(\alpha)\eta_0\Psi_3(\alpha^{-1}) \nn
 & & +\eta_0\Phi_4(-\alpha^{-1})\eta_0\Psi_1(-\alpha)
  Q_B\Phi_2(\alpha)\eta_0\Psi_3(\alpha^{-1})
\Big]\Bigg\rangle.
\eea
Again we reproduce the amplitude in the first quantization formalism:
\bea
A_{FBFB}
 & = &
  -\frac{1}{2}g^{-2}\int_0^1 d\alpha\frac{d\tau}{d\alpha}
  \Bigg\langle\int_{\bar{c}}\frac{dz}{2\pi i}\frac{dz}{dw}
  J^{--}(z)\Big[
 Q_B\Phi_4(-\alpha^{-1})\eta_0\Psi_1(-\alpha)
  \eta_0\Phi_2(\alpha)\eta_0\Psi_3(\alpha^{-1}) \nn
 & & +\eta_0\Phi_4(-\alpha^{-1})\eta_0\Psi_1(-\alpha)
  Q_B\Phi_2(\alpha)\eta_0\Psi_3(\alpha^{-1})
\Big]\Bigg\rangle \nn
 & = &
  -g^{-2}\int_0^1 d\alpha\frac{d\tau}{d\alpha}
  \Bigg\langle\int_{\bar{c}}\frac{dz}{2\pi i}\frac{dz}{dw}
  J^{--}(z)Q_B\Phi_4(-\alpha^{-1})\eta_0\Psi_1(-\alpha)
  \eta_0\Phi_2(\alpha)\eta_0\Psi_3(\alpha^{-1})\Bigg\rangle \nn
 & = &
  -g^{-2}\int_0^1 d\alpha
  \Bigg\langle\int d^2z\mu_\alpha(z,\bar{z})J^{--}(z)
  Q_B\Phi_4(-\alpha^{-1})\eta_0\Psi_1(-\alpha)
  \eta_0\Phi_2(\alpha)\eta_0\Psi_3(\alpha^{-1})\Bigg\rangle.
\eea
Thus all types of 4-point tree level amplitude coincide with
those of the first quantization formalism.

\section{Discussion}

Led by the analogy to type IIB supergravity, we have given a covariant
action for the fermion field with a constraint, and construct Feynman rules
for it. We have calculated on-shell tree level 4-point amplitudes with
fermions and have seen that they coincide with those of the first 
quantization formalism.

In our calculation of amplitudes we used only 3-point and 4-point vertices.
(In fact the 4-point vertices with fermions are absent. Therefore our 
calculation shows the correctness of their absence.) 
To confirm the correctness of higher vertices, we have to compute higher
correlators.

To compute loop amplitudes we need fermions, even if all the external 
legs are bosons, because fermions circulate through loops. 
Now that we have Feynman 
rules for fermions, in principle we can calculate any loop amplitude.
It is very interesting to see whether any cancellation expected from 
supersymmetry occurs among loop amplitudes. Another interesting issue
is to compute anomalies which come from fermion loops.

Pursuing the analogy to field theories with self-dual fields further,
it is natural to consider PST formalism-like formulation \cite{pst} 
i.e. introducing
auxiliary fields and constructing an action without any constraint.
It is intriguing to see if such formulation is possible in the open 
superstring field theory.

\vs{.5cm}
\noindent
{\large\bf Acknowledgments}\\[.2cm]
This work is supported by the Japan Society for the Promotion of Science 
and the Swiss National Science Foundation.

\newcommand{\J}[4]{{\sl #1} {\bf #2} (#3) #4}
\newcommand{\andJ}[3]{{\bf #1} (#2) #3}
\newcommand{\AP}{Ann.\ Phys.\ (N.Y.)}
\newcommand{\MPL}{Mod.\ Phys.\ Lett.}
\newcommand{\NP}{Nucl.\ Phys.}
\newcommand{\PL}{Phys.\ Lett.}
\newcommand{\PR}{Phys.\ Rev.}
\newcommand{\PRL}{Phys.\ Rev.\ Lett.}
\newcommand{\PTP}{Prog.\ Theor.\ Phys.}
\newcommand{\hepth}[1]{{\tt hep-th/#1}}

\end{document}